\documentclass[aps,prl,twocolumn,groupedaddress,nofootinbib]{revtex4-2}
\usepackage[utf8]{inputenc}
\usepackage[english]{babel}
\usepackage{amsmath}
\usepackage{graphicx}
\usepackage{dcolumn}
\usepackage{pbox}
\usepackage{amssymb}
\usepackage{epsfig}
\usepackage[dvipsnames]{xcolor}
\usepackage[normalem]{ulem}
\usepackage{slashed}
\usepackage{amssymb}
\usepackage{verbatim}
\usepackage{ mathrsfs }
\usepackage{color}
\usepackage{url}
\definecolor{MyDarkBlue}{rgb}{0.1, 0.1, 0.8}
\definecolor{SBlue}{rgb}{0.2, 0.4, 0.7} 
\definecolor{MyLightBlue}{rgb}{0.22,0.51,0.9}
\definecolor{MyGreen}{rgb}{0.0, 0.5, 0.0}
\definecolor{BrickRed}{rgb}{0.8, 0.25, 0.33}
\RequirePackage{hyperref}
\hypersetup{colorlinks, citecolor=SBlue,linkcolor=MyDarkBlue, urlcolor=PineGreen}

\begin{document}

\title{How common are grand unified theories?}

\author{Johannes Herms}
\email[]{johannes.herms@tum.de}
\affiliation{Max-Planck-Institut für Kernphysik, Saupfercheckweg 1, 69117 Heidelberg, Germany}

\author{Maximilian Ruhdorfer}
\email[]{m.ruhdorfer@cornell.edu}
\affiliation{Laboratory for Elementary Particle Physics, Cornell University, Ithaca, NY 14853, USA}


\date{\today}

\begin{abstract}
The individual fermion generations of the Standard Model fit neatly into a representation of a simple Grand Unified Theory gauge algebra. If Grand Unification is not realized in nature, this would appear to be a coincidence. We attempt to quantify how frequently this coincidence occurs among theories with group structure and fermion content similar to the Standard Model.
While many of the completely chiral, anomaly-free fermion representations of the Standard Model gauge algebra that are no larger than the single generation Standard Model are unifiable, we find that unifiability quickly becomes rare when the analysis is extended to include other gauge algebras or larger representations.
This purely group-theoretical analysis may be taken as a bottom-up indication for Grand Unification, conceptually similar to a naturalness argument.
\end{abstract}


\maketitle

The Standard Model (SM) of particle physics unifies electromagnetic and weak forces into a single framework~\cite{Glashow:1961tr,Weinberg:1967tq,Salam:1968rm}. The SM gauge forces, in turn, can be unified into a more symmetric Grand Unified Theory (GUT)~\cite{Pati:1974yy,Georgi:1974sy,Mohapatra:1974hk,Fritzsch:1974nn,Georgi:1974my}. Intriguingly, the SM fermions fit neatly into $SU(5)$ representations~\cite{Georgi:1974sy} and if a right-handed neutrino is added, one generation of fermions fits exactly into the $\mathbf{16}$ representation of $SO(10)$~\cite{Georgi:1974my,Fritzsch:1974nn}. This perfect fit seems to be too good to be a mere coincidence and is part of the appeal of GUTs. However, since symmetries and unification are driving concepts in physics, successfully constructing a GUT may be more of a result of our own preoccupations than an observation about nature.

In this work, we try to quantify how surprised we should be at the `unifiability' of the SM fermions.
We construct a base set of theories that look similar to the SM and check what fraction of them can be embedded in a GUT.
To obtain an answer, we need to define what we mean by \emph{`SM-like'} theories and which theories we consider to be \emph{`unifiable'}.
The result will depend on these arbitrary choices, but in a systematic way, allowing us to draw conservative conclusions.

\smallskip
\paragraph{Unifiability}

We use a UV-agnostic, bottom-up approach for unifiability where we ask if a given set of observed fermions by itself is unifiable into representations of a simple GUT algebra, without the need for additional, hitherto unobserved fermions.
The condition of no additional fermions 
provides closure to the problem and resembles the situation in the SM, where the known fermions unify into a representation of $SU(5)$.
Also note that we do not consider gauge coupling unification, as it is only suggestive in the SM and depends on the scalar sector as well. Our group-theoretic definition of unification is therefore only a necessary condition, such that our results will be conservative in the sense that the fraction of actually unifying theories will be smaller.
\begin{figure}[t]
    \centering
    \includegraphics[width=\columnwidth]{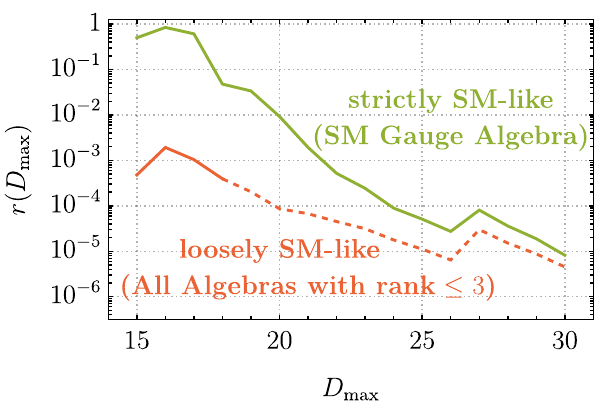}
    \caption{
    Fraction $r$ of SM-like, completely chiral, anomaly-free fermion representations that are unifiable into a representation of a simple GUT gauge algebra, as a function of the maximal considered fermion dimension $D_\mathrm{max}$. Representations are restricted to $U(1)$ charges of $|Q|<10$ and at most $\tilde{S}=4$ identical irreducible representations under the semi-simple part of the algebra.
    Green shows the result considering only representations of $SU(3)\times SU(2)\times U(1)$ (strictly SM-like theories), whereas red includes representations of all semi-simple gauge algebras with a $U(1)$ factor and rank smaller than three (loosely SM-like theories).
    For computational reasons, the number of anomaly-free fermion representations of $SU(2)^2\times U(1)$ and $SU(2)^3\times U(1)$ has been computed only up to $D_{\rm max} = 20$ and $D_{\rm max} = 18$, respectively. The dashed curve hence constitutes only an upper limit.
    }
    \label{fig:ratioPlot}
\end{figure}

\smallskip
\paragraph{Standard Model-like theories}
To assess the rarity of the unifiability property of the SM, we construct sets of theories that include and generalize the SM.
We consider the essential observational facts of the SM to be \emph{i)} three gauge forces corresponding to a reductive gauge algebra with a rank-3 semi-simple part, \emph{ii)} three generations of $D=15$ fermions each, with each generation anomaly free by itself, \emph{iii)} the fermions carry integer hypercharges $|Q|\leq 6$, \emph{iv)} the fermion representation is completely chiral.

\begin{figure*}[t]
    \begin{center}
    \includegraphics[width=\columnwidth]{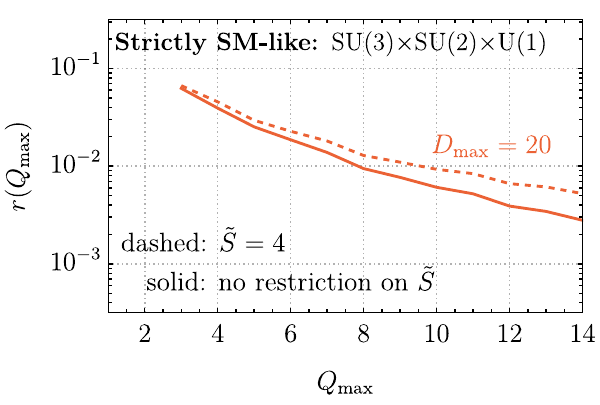}\hfill
    \includegraphics[width=\columnwidth]{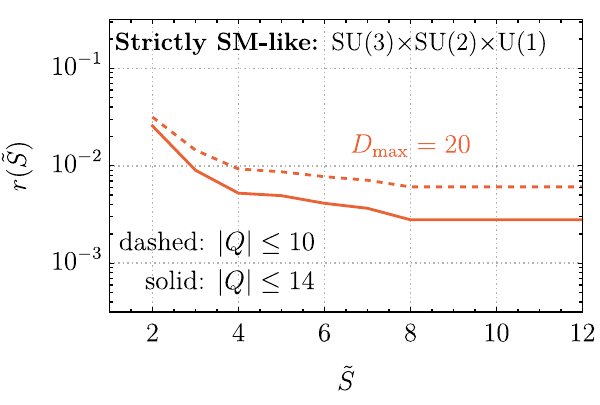}
    \end{center}
    \caption{
    Dependence of the unifiable fraction on $Q_{\rm max}$ and $\tilde S$ (strictly SM-like case).
    \emph{Left:} Cut on maximal considered integer charge $Q_{\rm max}$.
    \emph{Right:} Cut on the number $\tilde S$ of equal irreps of the semi-simple part of the gauge algebra.
    }
    \label{fig:QDS-dependence}
\end{figure*}

Based on these properties, we construct sets of \emph{SM-like} theories of increasing generality. Each theory is a completely chiral\footnote{We call a set of fermions 'completely chiral' if it contains no real (vector-like (VL)) subset.}, anomaly free
representation of a reductive gauge algebra consisting of a semi-simple part with rank $\leq 3$ and an abelian part with integer charges.
Concretely, we consider two definitions of SM-like:
\begin{enumerate}
\item {\it Strictly SM-like theories:} representations of the {\it SM} gauge algebra $SU(3)\times SU(2)\times U(1)$\,,
\item {\it Loosely SM-like theories:} representations of any semi-simple gauge algebra with rank $\leq 3$ and a $U(1)$ factor, i.e.  $SU(2)\times U(1)$ (rank-1), $\{SO(5), SU(2)^2, SP(4), G_2\}\times U(1)$ (rank-2) and $\{SU(2)^3,SU(4), SU(3)\times SU(2), SP(6),SO(5)\times SU(2), SO(7)\}\times U(1)$ (rank-3)\,.
\end{enumerate}
\noindent These base sets of self-consistent SM-like theories are made finite by imposing cuts on the maximal fermion dimension $D\leq D_\mathrm{max}$ and abelian charge $Q_\mathrm{max}$. 
We also restrict the number $\tilde S$ of identical irreducible representations (irreps) under the semi-simple part of the gauge algebra to $\tilde S \leq 4$ (everywhere but in Fig.~\ref{fig:QDS-dependence}) for computational reasons.
We determine the unifiable fraction of each set of SM-like theories as a function of $D_\mathrm{max}$ and $Q_\mathrm{max}$. The result depends on these arbitrary choices in a systematic way, which we will discuss below.
Note that our comparison is always with one generation of the SM where each generation unifies individually, i.e.\ we view the three-generation structure as a flavor symmetry which is also present in the GUT.

\smallskip
\paragraph{Likelihood of unifiability} We quantify the likelihood that a theory with an anomaly-free representation of dimension $D$ unifies into a simple GUT by the ratio of unifiable representations over all anomaly-free representations up to dimension $ D_{\rm max} \geq D$ for a given definition of SM-like. This is shown in Fig.~\ref{fig:ratioPlot} for both the strict and loose definition of SM-like.
The dependence of this likelihood on $D_{\rm max}$ will be discussed with our results.

\section{Methods}
There are two steps to assessing how common unifiability is among SM-like theories.
In a bottom up approach, we first construct a base set of all consistent (i.e. anomaly-free), completely-chiral SM-like theories.
Then, we check unifiability for each of them, using the SuperFlocci~\cite{Gomes:2023pwj} code.
Using the GroupMath~\cite{Fonseca:2020vke} code, we can verify and extend our results by a top-down determination of branchings of all candidate GUTs.

\smallskip
\paragraph{Constructing anomaly-free representations}
We construct the set of anomaly-free representations of a given gauge algebra and values of $D_\mathrm{max}$, $Q_\mathrm{max}$ in three steps:
\begin{enumerate}
    \item [i)] Find all anomaly-free representations of the semi-simple part of the gauge algebra (i.e.\ $SU(3)\times SU(2)$ in the SM) up to dimension $D_{\rm max}$.
    \item [ii)] Assign integer $U(1)$ charges within a predefined range $|Q| \leq Q_{\rm max}$ to all representations under the semi-simple part of the algebra and keep those that are completely chiral and satisfy the anomaly cancellation conditions (gravitational and gauge anomalies).\footnote{Note that since we work at the level of Lie algebras we do not check for global anomalies, such as the anomaly associated with an odd number of fermion doublets charged under $SU(2)$ in four dimensions~\cite{Witten:1982fp}. Such global anomalies depend on the global structure of the Lie group and cannot be determined from the Lie algebra alone as there is no one-to-one correspondence between Lie groups and algebras. However, we have checked that excluding representations with an odd number of $SU(2)$ doublets in the base set changes our results by an $\mathcal{O}(1)$ number (e.g. the result in eq.~(\ref{eqn:conservativeres}) does not change, while eq.~(\ref{eqn:naiveres}) changes to $11/496$).}
    \item [iii)] Filter out equivalent representations. We consider representations equivalent if they differ only by an integer rescaling of the $U(1)$ charge or are conjugate representations of each other (or a combination of both). For this reason we keep only representations where the greatest common divisor of all charges is one.
    An example of equivalent $(SU(3),SU(2))_{U(1)}$ representations is
    \begin{equation*}
    \begin{split}
        &(\mathbf{3},\mathbf{2})_0 \oplus (\mathbf{\bar{3}},\mathbf{1})_{-1} \oplus (\mathbf{\bar{3}},\mathbf{1})_{1} \\
        &\qquad\qquad\quad\quad \sim (\mathbf{\bar{3}},\mathbf{2})_0 \oplus (\mathbf{3},\mathbf{1})_{-2} \oplus (\mathbf{3},\mathbf{1})_{2}\,,
    \end{split}
    \end{equation*}
    since they are conjugate representations with rescaled $U(1)$ charges.
\end{enumerate}
While i) and iii) are easily implemented with Mathematica packages such as SuperFlocci~\cite{Gomes:2023pwj} or GroupMath~\cite{Fonseca:2020vke}, ii) is a challenging combinatoric problem since the number of possible charge assignments grows exponentially with the number of fermions. In order to deal with this large number of possible charge assignments we use compiled Mathematica code and simplify the problem for a given semi-simple representation in the following way. We generate charge assignments for blocks of identical irreps within a given candidate representation in a lexicographic order that does not go through permutations of charges within one block. Additionally we split the candidate representation in two and compute anomaly coefficients for charge assignments in each part separately. Finally we match those assignments which add up to zero when combined from the two halves.
Despite the above simplifications we cannot handle semi-simple representations which are composed of a large number of irreps. For this reason we restrict ourselves to representations that contain no more than $\tilde S \leq 4$ equal semi-simple irreps.
This is on the one hand necessary to limit the number of $U(1)$ charge assignments when extending the analysis to large fermion dimensions and on the other hand mimics the situation in the SM where no semi-simple irrep appears more than twice in a generation.
We further demand that all forces have at least one particle charged under them.

\smallskip
\paragraph{SuperFlocci~\cite{Gomes:2023pwj} -- bottom-up determination of unifiability}
SuperFlocci is a Mathematica package that ``takes any reductive gauge algebra and fully-reducible fermion representation, and outputs all semi-simple gauge extensions under the condition that they have no additional fermions, and are free from local anomalies''~\cite{Gomes:2023pwj}.
A theory is unifiable according to the definition laid out in the introduction exactly if SuperFlocci finds a simple gauge extension. 

\smallskip
\paragraph{GroupMath~\cite{Fonseca:2020vke} -- top-down construction of unifiable representations}
As a second approach, we consider all candidate unified (i.e.\ simple) gauge algebras that have (non-singlet) representations with dimension $\leq D_\mathrm{max}$. We use the GroupMath Mathematica package to find all decompositions of all candidate GUT representations to the reductive gauge algebra under consideration.
We assign $U(1)$-charges according to the rules determined by GroupMath and apply the same filters to the resulting representations as to the base set.
This is an independent top-down check on the number of unifiable theories and gives the same result as the bottom-up analysis using SuperFlocci.
It is, however, more efficient and allows to extend the analysis to larger fermion dimensions.

\smallskip

\paragraph{Examples}
To illustrate our approach we provide a few examples for consistent theories that we do or do not consider SM-like according to our strict definition, and that are or are not unifiable.

The smallest fermion representation free from local anomalies 
under the SM gauge algebra $\mathrm{SU}(3)\times \mathrm{SU}(2)\times \mathrm{U}(1)$, which has particles charged under all three forces, is:
\begin{equation}
(\mathbf{1}, \mathbf{2})_0 \oplus 
(\mathbf{3}, \mathbf{1})_{-1} \oplus 
(\bar{\mathbf{3}}, \mathbf{1})_1 \,.
\end{equation}
It unifies into an $\mathbf{8}$ of $\mathrm{Sp}(8)$, but is vector-like and therefore not included in the base set.

The first completely chiral representation appears at $D=12$: 
\begin{equation}
(\mathbf{3}, \mathbf{2})_0 \oplus 
(\bar{\mathbf{3}}, \mathbf{1})_{-1} \oplus 
(\bar{\mathbf{3}}, \mathbf{1})_1 \,.
\label{eqn:smallestChiral}
\end{equation}
It is not unifiable.
The smallest completely chiral, unifiable representation of the SM gauge algebra is the single-generation SM at $D=15$:
\begin{equation}
(\mathbf{1}, \mathbf{1})_{-6} \oplus
(\mathbf{1}, \mathbf{2})_3 \oplus 
(\bar{\mathbf{3}}, \mathbf{2})_{-1} \oplus 
(\mathbf{3}, \mathbf{1})_{-2} \oplus 
(\mathbf{3}, \mathbf{1})_4 \,.
\label{eqn:SMrep}
\end{equation}

\section{Results}
With the techniques described above, we can assess how common the property of unifiability is among low-dimensional representations of SM-like gauge algebras.
To this end, we progressively generalize the scope of the SM-like base set.

\smallskip
\paragraph{Strictly SM-like}
To begin with, we consider \emph{strictly SM-like} theories, i.e.\ representations of the SM algebra only.
If we restrict ourselves to completely chiral representations no larger than the single-generation SM itself (i.e.\ $D_\mathrm{max}=15$), the only anomaly free representations are those given in (\ref{eqn:smallestChiral}) and (\ref{eqn:SMrep}), resulting in the unifiable fraction
\begin{equation}
    \left.\frac{\# \text{ unifiable reps}}{\# \text{ strictly SM-like reps}}\right|_{D_\mathrm{max}=15}^{\text{completely chiral}}
     =
     \frac{1}{2} \,,
     \label{eqn:conservativeres}
\end{equation}
which can also be seen from the green curve at $D_{\rm max} =15$ in Fig.~\ref{fig:ratioPlot}.
To make statements about the prevalence of any property, we clearly need to broaden the definition of SM-like.

Fig.~\ref{fig:ratioPlot} extends this result and shows the ratio $r$ of the number of completely chiral unifiable representations over the number of all completely chiral SM-like fermion representations of the SM gauge algebra, depending on the fermion dimension $D \leq D_\mathrm{max}$ and restricted to representations with at most $\tilde{S}\leq 4$ identical irreps under the semi-simple part of the algebra and $U(1)$ charges $|Q|\leq 10$.
Picking for example $D_\mathrm{max}=20$ as arbitrary cut on what is considered to be SM-like, we have (see Fig.~\ref{fig:ratioPlot})
\begin{equation}
        \left.\frac{\# \text{ unifiable reps}}{\# \text{ strictly SM-like reps}}\right|_{D_\mathrm{max}=20,\;|Q|\leq 10}^{\text{completely chiral}}
     =
     \frac{11}{1186}
     \,.
     \label{eqn:naiveres}
\end{equation}
This number depends sensitively on the arbitrary choices of $D_\mathrm{max}$ and $Q_\mathrm{max}$. 
We also find that if we do not restrict $\tilde{S}$, the base set is inflated by theories with a large number of semi-simple singlets or $SU(2)$ doublets for larger $D_{\rm max}$.
Figs.~\ref{fig:ratioPlot} and~\ref{fig:QDS-dependence} show how the result depends on these arbitrary cuts.
For all of $Q_\mathrm{max}$, $D_\mathrm{max}$ and $\tilde S$, there is a clear trend of falling unifiable fraction for more inclusive definitions of SM-like. As can be seen in both panels of Fig.~\ref{fig:QDS-dependence}, more restrictive cuts lead to more conservative estimates, i.e.\ larger values of $r$.

\begin{figure*}[tb]
    \centering
    \includegraphics[width=\columnwidth]{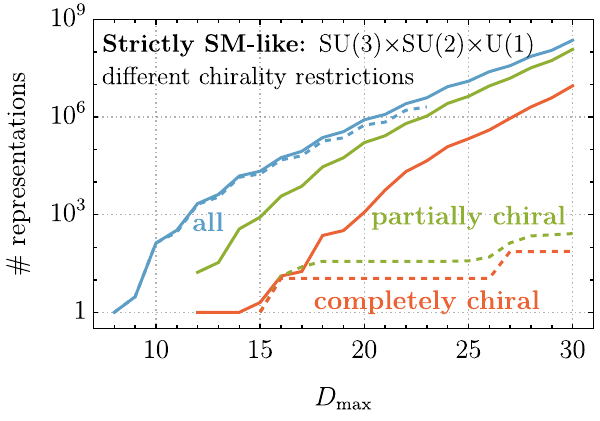}\hfill
    \includegraphics[width=\columnwidth]{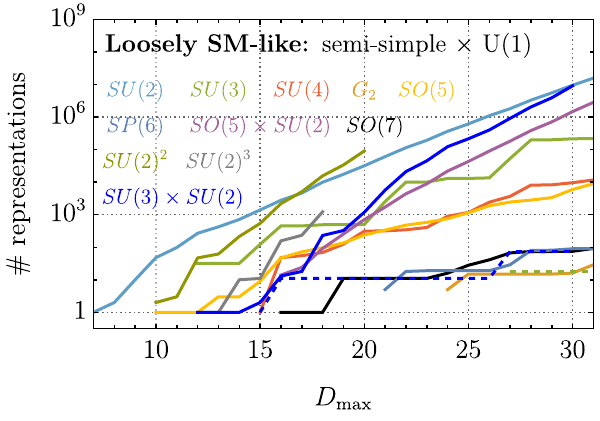}
    \caption{
    Number of anomaly-free fermion representations up to dimension $D_\mathrm{max}$. 
    Solid lines correspond to all anomaly-free representations, while dashed lines only count those that unify.
    $Q_\mathrm{max}$ is fixed to $10$ in this figure.
    \emph{Left:} Strictly SM-like theories (ie.\ SM gauge algebra representations), showing relaxed chirality restrictions. The red lines show completely chiral representations (fiducial case). The green lines show results also including partially chiral representations, while the blue lines additionally include completely VL representations.
    The jumps in the number of unifiable theories occur when unification into $SO(10)$ ($SU(6)$ and $E_6$) become possible at $D_{\rm max}=16$ ($D_{\rm max} =27$).
    \emph{Right:} Contribution from the different gauge algebras to the number of loosely SM-like theories. 
    Due to computational reasons we only determine the number of representations for $SU(2)^2\times U(1)$ and $SU(2)^3\times U(1)$ up to $D_{\rm max}=20$ and $D_{\rm max} = 18$, respectively.
    }
    \label{fig:numRep}
\end{figure*}

\smallskip
\paragraph{Loosely SM-like}
Next, we generalize the analysis to \emph{loosely SM-like} theories, i.e.\ representations of 
all reductive semi-simple $\times$ abelian gauge algebras with rank of the semi-simple algebra $\leq 3$, while keeping the requirements $|Q| \leq 10$ and $\tilde S \leq 4$.
The corresponding unifiable fraction is shown in Fig.~\ref{fig:ratioPlot}, and the number of anomaly-free representations for each of the semi-simple algebras is shown in the right panel of Fig.~\ref{fig:numRep}.\footnote{Note that due to computational reasons we did not determine the number of anomaly-free representations in the base set of $SU(2)^2\times U(1)$ and $SU(2)^3\times U(1)$ beyond $D_{\rm max}=20$ and $D_{\rm max}=18$, respectively. For this reason the red curve is dashed in Fig.~\ref{fig:ratioPlot} for $D_{\rm max} > 18$. The dashed curve can be viewed as a conservative estimate for $r(D_{\rm max})$ with the real value, including all representations, being smaller.}
We find that among completely chiral representations of SM-like gauge algebras with dimension and charges no larger than that of the SM,
\begin{equation}
    \left.\frac{\# \text{ unifiable reps}}{\# \text{ loosely SM-like reps}}\right|_{D_\mathrm{max}=15,\;|Q|\leq 6}^{\text{completely chiral}}
     =
     \frac{1}{365}
    \label{eqn:conservativeAllAlgebras}
\end{equation}
are unifiable
(see also red curve in Fig.~\ref{fig:ratioPlot} for $|Q|\leq10$).
In the case of $SU(3)\times U(1)$, we need to go to $D \geq 27$ to find the first completely chiral, unifiable representations (which unify into $SU(6)$ and $E_6$ representations). 
For the remaining rank-1 ($SU(2)\times U(1)$) and rank-2 gauge algebras $\{SO(5),SU(2)\times SU(2), SP(4), G_2\}\times U(1)$ we do not find any chiral, unifiable fermion representations with $D\leq 30$. The same is true for rank-3 algebras $\{SU(2)^3,SU(4), SP(6),SO(5)\times SU(2), SO(7)\}\times U(1)$, with the exception of the SM gauge algebra. This is partially due to the increasing dimension of the smallest representations of these algebras. At the same time the base set of completely chiral theories is still growing exponentially with $D$ since $U(1)$ charge assignments can be used to make sets of five or more fermions chiral~\cite{Batra:2005rh,Costa:2019zzy}.

\paragraph{The role of chirality}
Instead of considering completely chiral fermion representations, we can extend the analysis to also include \emph{partially chiral} representations (including singlets or VL particles), as shown in Fig.~\ref{fig:numRep} for the case of the SM gauge algebra (strictly SM-like).
There are many more anomaly-free partially chiral representations than completely chiral ones, as can be understood from the large number of possible charge assignments for VL-pairs in a representation.
The number of unifiable representations also increases, but much less dramatically.
Compared to the completely chiral case in eqns.~(\ref{eqn:conservativeres}),(\ref{eqn:naiveres}), when partially chiral representations are included, the most restrictive SM neighborhood ($D_\mathrm{max}=15,\;|Q|<6,\;\tilde{S} \leq 4$, strictly SM-like)  yields $r = 1/111$, and for $D_\mathrm{max} = 20,\; |Q|\leq 10,\;\tilde{S} \leq 4$, we find $r\sim 2 \cdot 10^{-4}$.
Relaxing all chirality restrictions is not very informative, as both the base set and the unifiable set are then dominated by the large number of completely VL representations (see blue curve in the left panel of Fig.~\ref{fig:numRep}), which are very unlike the SM.
The number of theories quickly becomes difficult to handle computationally, but for $D_\mathrm{max} \leq 20,\, |Q|\leq 10,\, \tilde S \leq 4$ we find $r \sim 1$, i.e.\ most completely VL representations with small integer charges can be obtained when decomposing a VL representation of some unified gauge algebra. The relative rarity of unifiable representations in the chirality restricted cases considered above then can be interpreted as resulting from the small number of non-VL candidate GUT representations.

\section{Discussion}
We now return to the question we set out to answer, namely whether the fact that the observed fermions of the SM can be unified is in itself surprising and may be a hint to what lies beyond the SM, or just a quirk of group theory common among representations of reductive Lie algebras.
Different definitions of the base set of SM-like theories result in different values of the unifiable fraction, with a clear pattern.
Many of the completely chiral representations of the SM gauge algebra (ie.\ strictly SM-like theories) with dimension and charges not much larger than the single generation SM 
($D_\mathrm{max}\lesssim 17,\;|Q|\leq 10,\; \tilde{S} \leq 4$)
can be exactly embedded into a representation of $SU(5)$ or $SO(10)$, resulting in an $\mathcal{O}(1)$ unifiable fraction.
However, once we look at any more general set of consistent fermion representations beyond this immediate neighborhood of the single generation SM, unifiability becomes a rare property:
When allowing the dimension of the representation to be only a little larger than that of the SM, unifiability quickly becomes rarer than $10^{-2}$ (see Fig.~\ref{fig:ratioPlot}).
Including partially chiral representations also pushes the unifiable fraction below $10^{-2}$ (see Fig.~\ref{fig:numRep}).
Considering all small reductive gauge algebras with an abelian part (ie.\ loosely SM-like theories), only $10^{-3}$ of the minimal neighborhood (ie.\ representations with $D_\mathrm{max}\leq 15,\;|Q|\leq 6,\; \tilde{S} \leq 4$) of the single-generation SM is unifiable (see eqn.~\ref{eqn:conservativeAllAlgebras}).
This demonstrates that fermion unifiability as it exists generation-by-generation in the Standard Model is not a common property among similar quantum field theories.

Let us also mention that different definitions of unifiability are possible.
Our approach of demanding that the fermions fit neatly into a representation of the GUT that includes no further fermions may be relaxed.
From a practical perspective, there is no harm if a GUT predicts hitherto unobserved VL fermions -- those may be heavy.
This is the case for instance for the right handed neutrinos predicted in $SO(10)$ grand unification~\cite{Fritzsch:1974nn}.
A new condition to provide closure to the problem would be needed (eg.\ an arbitrary cut on the number of inferred VL fermions).
Using our existing results, we can estimate the impact of relaxing the unifiability criterion by considering all theories as unifiable that unify when adding VL fermions up to a total fermion dimension of $30$: In this case, the unifiable fraction of SM representations in Fig.~\ref{fig:ratioPlot} increases by a factor of up to 2, and the first partially chiral unifiable theory appears at $D = 12$.
We leave the study of stronger unification criteria (eg.\ unification of fermions into an irreducible representation, as in $SO(10)$ GUTs) to future work.

\section{Conclusion}
In this work, we consider the single-generation SM as one among many similar consistent theories.
From this starting point, the observation that the SM fermion representation is unifiable may seem surprising. Here we try to find a quantitative answer to the question of how surprised we should actually be.
We find that the unifiability of fermions into a representation of a simple unified algebra, as it occurs in the SM into a $SU(5)$ GUT, is rare ($< 10^{-2}$) among SM-like chiral theories once we go beyond the immediate neighborhood of the single-generation SM. 
The argument presented here can be taken as a purely group-theoretical indication for Grand Unification, conceptually similar to a naturalness argument. 
However, the absence of a probability measure in the space of theories hampers a probabilistic interpretation of our results.

\begin{acknowledgments}
 \emph{We thank}
 Joseph Tooby-Smith for collaboration in the early stages of this project and coding support throughout. We also thank Andreas Trautner and Joseph Tooby-Smith for valuable comments on the manuscript. MR is supported by the NSF grant PHY-2014071. The work of MR was performed in part at the Aspen Center for Physics, which is supported by a grant from the Simons Foundation (1161654, Troyer).
\end{acknowledgments}


\bibliography{pGUT}

\end{document}